# A general framework for reversible data hiding in encrypted images by reserving room before encryption

Ammar Mohammadi

*Abstract*— In this paper a general framework to adopt different predictors for reversible data hiding in the encrypted image is presented. We propose innovative predictors that contribute more significantly than conventional ones results in accomplishing more payload. Reserving room before encryption (RRBE) is designated in the proposed scheme making possible to attain high embedding capacity. In RRBE procedure, pre-processing is allowed before image encryption. In our scheme, pre-processing comprises of three main steps: computing prediction-errors, blocking and labeling of the errors. By blocking, we obviate the need for lossless compression to when content-owner is not enthusiastic. Lossless compression is employed in recent state of the art schemes to improve payload. We surpass prior arts exploiting proper predictors, more efficient labeling procedure and blocking of the prediction-errors.

## I. INTRODUCTION

In some applications of data embedment in an image it is needed to reconstruct the original image without any distortion. To satisfy this requirement, reversible data hiding (RDH) is an approach employed in some applications to satisfy complete restoration of the original image after error-free extraction of the embedded data. The most presented schemes in RDH inspire from three main notions, namely, histogram modification, difference expansion, and lossless compression pioneered by [1], [2] and [3], respectively.

In an extension, RDH in the encrypted image has drastically attracted research interest, due to the growth of cloud computing that is reliable, available and efficient in cost. To keep privacy in using cloud computing, content-owner is motivated to encrypt his/her sensitive data including private images. Data hider, as network administrator or cloud provider, needs to embed some notifications in the encrypted image. At the recipient side, authorized users should be able to extract embedded data and restore the original image. Therefore, lossless original image reconstruction and error-free data extraction should be guaranteed the cause of developing RDHEI.

Generally, RDHEI schemes are classified into three groups: reserving room before encryption (RRBE) [4-15], vacating room by encryption (VRBE) [16-22] and vacating room after encryption (VRAE) [23-29].

In VRAE schemes, data hider should designate a procedure to embed data in the encrypted image without having any knowledge of the original content. Therefore, in VRAE, data hider is completely blind to the spatial correlation of the original content. At the recipient side, process of extracting data may be accomplished jointly or separately with the process of reconstructing the original image. In comparison with a joint one, a separable procedure of VRAE is more practical and correspondingly more challengeable in achieving high data embedment and realizing lossless reconstruction. Compression of the encrypted image [26, 28] and MSB integration of the encrypted pixels, [24], are two substantial procedures realizing separable VRAE. Note that, the idea of compressing the encrypted information is introduced in [30, 31].

In the most procedures to realize VRBE, all pixels in a local area of the image are encrypted by a similar cipher byte. Therefore, spatial correlation of the pixels are partially preserved to provide room for embedding data bits. By preserving correlation, Huang *et al.*, [17] present a new framework through VRBE that makes possible to employ methods of RDH in RDHEI.

Different procedure than VRAE and VRBE, in RRBE, spatial correlation may be completely used to achieve higher embedding capacity by applying pre-processing before encryption. Note that, almost all presented schemes in RDHEI employ correlation of the neighboring pixels. In older methods of RRBE, pre-processing is a self-reversible embedding procedure. The schemes of Cao *et al.* [4] and Ma *et al.* [6] use the self-reversible embedding to accommodate some areas of the original image in the smoother areas through traditional methods of RDH. Employing Pailier cryptosystem for encryption and self-reversible embedding, Xiang and Luo [14] accomplish a pre-processing to release room for embedding data bits. In the most schemes of RRBE, the important portion of pre-processing is to compute prediction-errors. In the prediction of a pixel, the more correlated the neighboring pixels are the smaller the prediction-error will be. Meanwhile, different predictors may be exploited to compute prediction-errors. The better the predictor is the sharper the histogram of the errors obtains and correspondingly the higher the room will provide. In recent years, causal predictors MED, [42], and GAP, [43], and accordingly, non-causal predictors chess-board (CB), [44], and local difference (LD), [45], are used by many schemes of RDH and RDHEI as listed in Table I. As described, more developed schemes in RDH and RDHEI employ CB and MED predictors, respectively. Although CB predictor is more efficient than others (refer to Subsection IV-A), it is a non-causal predictor and cannot be appropriate in some reported techniques of RDHEI. In Table I, six out of nine listed schemes of RDHEI use RRBE. In the pre-processing, prediction-errors may be computed by the nominated predictors. By lossless compression of the prediction-errors, Yin *et al.* [12] conduct a procedure to release room. In different procedure, by analyzing these errors, some information or labels of the original image may be extracted and employed to embed secret data after encryption [7-10]. These labels, as overhead data, show the level of pixels smoothness in the original image exploited to embed data efficiently in the encrypted image. Along with secret data, overhead data is embedded in the encrypted image to be used by the authorized recipients for data extraction and original image reconstruction. In scheme [9], overhead data losslessly are compressed by entropy coding to provide more pure embedding capacity. However, in some applications, content-owner is not enthusiastic to employ lossless compression. In [7], by dividing an image to non-overlapped blocks and attaining prediction-errors in each block using LD predictor, we improve RRBE schemes without using lossless compression. In this paper, we extend scheme of [7] by introducing a general framework to employ more effective predictors that significantly exceeds the payload.

TABLE I CONTRIBUTION OF CONVENTIONAL PREDICTORS IN RDH AND RDHEI SCHEMES

| Predictors | RDH | RDHEI |
|---|---|---|
| **MED** | [32, 33] | [9-12] |
| **GAP** | [34, 35] | |
| **CB** | [36-41] | [6, 24, 25] |
| **LD** | | [7, 19] |

The main contribution of this paper may be summarized as follows:

1. A general framework without using lossless compression is proposed that makes possible to employ different predictors. Therefore, advantage of spatial correlation can be fully taken by adopting the best possible predictor. It results in releasing more room to convey secret data that improves embedding capacity.
2. Novel predictors, L and adaptive L, are proposed to develop not only the proposed scheme but also the prior arts of RDHEI and RDH.
3. At the recipient side, lossless reconstruction of the original image and error-free extraction of the secret data are done just by having content-owner key and data hider key, respectively.

The rest of the paper is organized as follows. Different predictors including proposed ones are demonstrated in Section II that may be employed in our general framework. The general frame work is presented in Section III. Section IV shows the experimental results and Section V concludes the paper.

## II. PROPOSED PREDICTORS

In this section, we describe conventional predictors, namely, GAP, MED and CB. By modifying GAP predictor we present simple GAP (SGAP) predictor. Through improving SGAP, we propose L and adaptive L predictors.

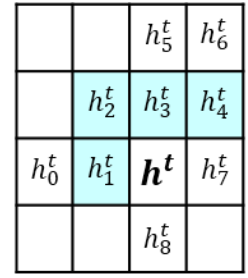

Fig. 1. A neighborhood of the target pixel, $h^t$.

$$\underbrace{\begin{bmatrix} h_1^1 & h_2^1 & h_3^1 & h_4^1 \\ \vdots & \vdots & \vdots & \vdots \\ h_1^t & h_2^t & h_3^t & h_4^t \\ \vdots & \vdots & \vdots & \vdots \\ h_1^T & h_2^T & h_3^T & h_4^T \end{bmatrix}_{T\times 4}}_{\mathbf{X}} \underbrace{\begin{bmatrix} a_1 \\ a_2 \\ a_3 \\ a_4 \end{bmatrix}_{4\times 1}}_{\mathbf{A}} = \underbrace{\begin{bmatrix} h^1 \\ \vdots \\ h^t \\ \vdots \\ h^T \end{bmatrix}_{T\times 1}}_{\mathbf{Y}}$$

Fig. 2. Linear regression model.

### A. Conventional predictors

As shown in Fig. 1, suppose that $h^t$ is the target pixel that may be predicted by its neighboring pixels, $\{h_1^t, ..., h_8^t\}$, and $\tilde{h}^t$ is the predicted value of $h^t$. In CB predictor, $\tilde{h}^t$ is achieved by

$$\tilde{h}^t = (h_1^t + h_3^t + h_7^t + h_8^t)/4 \qquad (1)$$

using Med predictor, $\tilde{h}^t$ is calculated according to

$$\tilde{h}^t = \begin{cases} \max(h_1^t, h_3^t) & \text{if } h_2 \leq \min(h_1^t, h_3^t) \\ \min(h_1^t, h_3^t) & \text{if } h_2 \geq \max(h_1^t, h_3^t) \\ h_1^t + h_3^t - h_2^t & \text{else} \end{cases} \qquad (2)$$

and in GAP, prediction is given by

$$\tilde{h}^t = \begin{cases} h_1^t & \text{if } (d_v - d_h) > 80 \\ (h_1^t + u)/2 & \text{if } (d_v - d_h) \in (32,80] \\ (h_1^t + 3u)/4 & \text{if } (d_v - d_h) \in (8,32] \\ u & \text{if } (d_v - d_h) \in [-8,8] \\ (h_3^t + 3u)/4 & \text{if } (d_v - d_h) \in [-32,-8) \\ (h_3^t + u)/2 & \text{if } (d_v - d_h) \in [-80,-32) \\ h_3^t & \text{if } (d_v - d_h) < -80 \end{cases}$$
$d_v = |h_1^t - h_2^t| + |h_4^t - h_6^t| + |h_3^t - h_5^t|, d_h = |h_1^t - h_0^t| + |h_4^t - h_3^t| + |h_3^t - h_2^t|, u = (h_1^t + h_3^t)/2 + (h_4^t - h_2^t)/4.$
(3)

Discarding $d_v$ and $d_h$, prediction may be done just using the term $u$ as follows:

$$\tilde{h}^t = u = 0.5h_1^t - 0.25h_2^t + 0.5h_3^t + 0.25h_4^t \qquad (4)$$

In the closed form, it can be a matrix multiplication of coefficient matrix, $\mathbf{A}' = [0.5, -0.25, 0.5, 0.25]$, and target matrix, $\mathbf{H_t} = \{h_1^t, h_2^t, h_3^t, h_4^t\}$ as follows:

$$\tilde{h}^t = \mathbf{H_t} \times \mathbf{A} \qquad (5)$$

which introduces a simple GAP that is slightly less efficient than GAP but much less complex. In this context, $(.)'$ is the transpose operation.

*B. L predictor*

Adopting more efficient coefficient for SGAP predictor, $\mathbf{A}' = [0.7, -0.3, 0.5, 0.1]$, we present L predictor is given by

$$\tilde{h}^t = \mathbf{H_t} \times \mathbf{A} = 0.7h_1^t - 0.3h_2^t + 0.5h_3^t + 0.1h_4^t \qquad (6)$$

It can be more effectual, if the coefficient of L predictor is adaptively calculated for a specific image. It will be realized by introducing adaptive L (AL) predictor discussed in the next Subsection. The coefficient of L predictor, in effect, is attained by average of thousands determined coefficients computed by AL predictor for various images.

*C. Adaptive L (AL) predictor*

No matter what kind of image you choose, a determined coefficient is employed to predict pixels through L predictor; however various images may have different statistical features. In more efficient procedure, the coefficient can be computed adaptively in proportion to each image or separated blocks of an image. Let's have the number of $T$ different target pixels in a block, where $h^t$ and $\{h_1^t, h_2^t, h_3^t, h_4^t\}$, $1 \leq t \leq T$, respectively, are intensities of the target pixel and it's neighboring pixels. As shown in Fig. 2, these target pixels and their four neighbors describe in the form of matrixes $\mathbf{Y}$ and $\mathbf{X}$, respectively. The goal is to attain more efficient coefficient, $\mathbf{A}' = [a_1, a_2, a_3, a_4]$, that minimizes sum of squared errors, $\mathcal{E} = \|\mathbf{Y} - \mathbf{XA}\|^2$. In effect, finding the efficient coefficient is a linear regression problem, [46], that fits a line to data and it may be achieved by regarding the least-squares estimation as follows

$$\arg\min_{\mathbf{A}}(\mathcal{E}(\mathbf{A})) = \operatorname*{argmin}_{\mathbf{A}}(\|\mathbf{Y} - \mathbf{XA}\|^2) \qquad (7)$$

where $\| \ \|$ is Euclidean norm and $\mathcal{E}$ is a function of $\mathbf{A}$ can be rewritten as

$$\mathcal{E}(\mathbf{A}) = \|\mathbf{Y} - \mathbf{XA}\|^2 = (\mathbf{Y} - \mathbf{XA})'(\mathbf{Y} - \mathbf{XA}) = \mathbf{Y}'\mathbf{Y} - \mathbf{Y}'\mathbf{XA} - \mathbf{A}'\mathbf{X}'\mathbf{Y} - \mathbf{A}'\mathbf{X}'\mathbf{XA} \qquad (8)$$

As $\mathcal{E}(\mathbf{A})$ is convex, optimum value of $\mathbf{A}$ is achieved by setting the gradient to zero.

$$\frac{\partial(\mathcal{E}(\mathbf{A}))}{\partial \mathbf{A}} = -2\mathbf{Y}'\mathbf{X} - 2\mathbf{A}'\mathbf{X}'\mathbf{X} = 0 \qquad (9)$$

As a result

$$\mathbf{A} = (\mathbf{X}'\mathbf{X})^{-1}\mathbf{X}'\mathbf{Y} \qquad (10)$$

### III. PROPOSED FRAMEWORK

In our previous work [7], a method in RDHEI is presented that employs a non-causal predictor, LD, to compute prediction-error (PE). In LD predictor, in a local area of the image, the pixels are predicted by the most central pixel denoted reference pixel. It remains intact during embedding process. In this paper, we conduct the scheme of [7] in a general form to exploit different causal predictors that are more efficient than LD. Accordingly, all pixels of the image can be used for embedding data results in improving payload. Fig. 3 shows a generic block diagram of the proposed method. Fig. 3a illustrates the procedure of embedding secret data in the encrypted image that results in the marked image and Fig. 3b demonstrates original image recovery and secret data extraction. Embedding procedure comprises pre-processing, encryption and embedding secret data. The pre-processing itself consists of computing prediction-errors, blocking the errors and attaining labels. In prediction, an appropriate causal predictor may be chosen. Different causal predictors including proposed predictors are demonstrated in Section II that may be exploited in the proposed framework. As discussed, in [7], we use LD predictor to provide room. Definitely using more efficient predictors in a general framework, we may outperform our previous work.

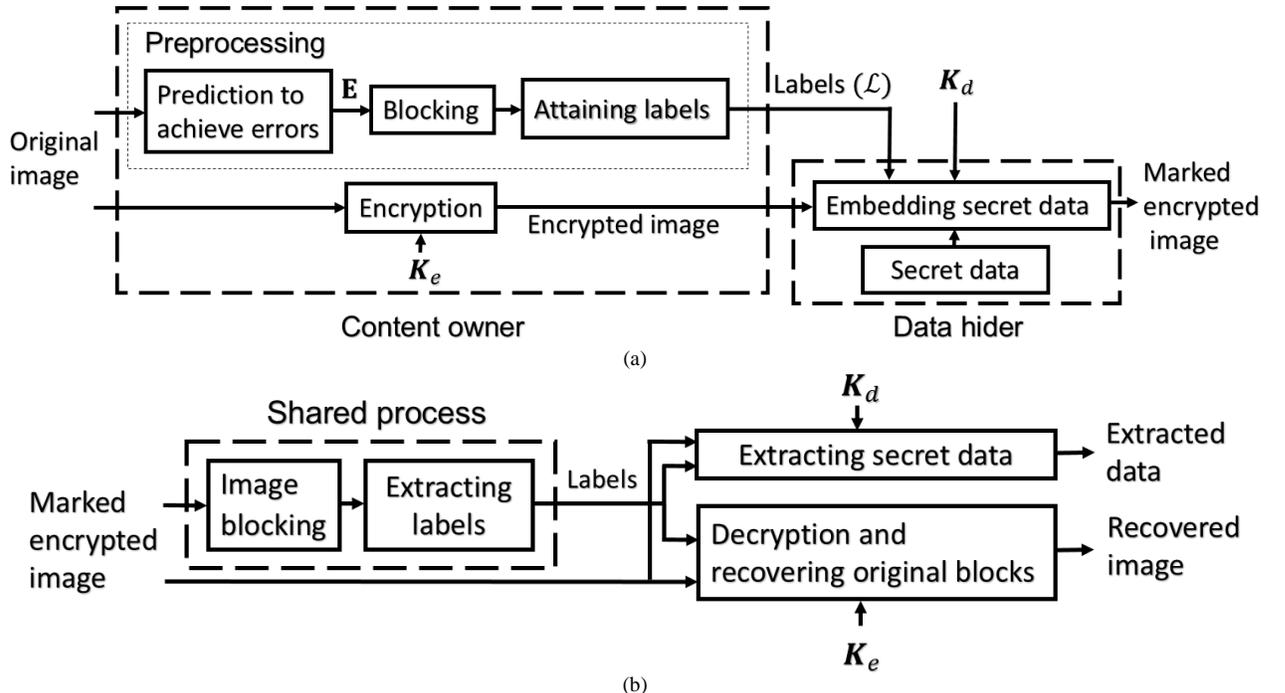

Fig. 3. Generic block diagram of the proposed scheme. (a) Embedding data and forming marked encrypted image. (b) Extracting data and reconstructing the original image.

At the recipient side, the first step is to bring out the labels. Having labels, secret data can be completely extracted and original image may be losslessly reconstructed. In the following, procedures of data embedding and extracting are explained in four subsections.

### A. Preprocessing

Preprocessing is initiated by achieving prediction-errors (PEs) using a causal predictor such as MED, GAP, L and AL described in Section II. For pixels of the original image, corresponding PEs are computed, and denoted by $\mathbf{E}$. Then, $\mathbf{E}$ is divided into non-overlapped blocks. Afterwards, regarding the maximum PE of each block, a label is devoted that indicates embedding capacity of the block. We divide $\mathbf{E}$ to $K$ blocks with the size of $q \times w$. In the block $k$, $1 \leq k \leq K$, $e^{(k,i)}$ is $i$'th PE designated for the corresponding target pixel $h^{(k,i)}$ of the host image, ($1 \leq i \leq I = q \times w$). Data is embedded in the target pixel. Adopting $t = (k-1) \times I + i$, $h^t$ and $e^t$ are a compact notification of $h^{(k,i)}$ and $e^{(k,i)}$ so prediction-error is obtained by

$$e^t = h^t - \tilde{h}^t \quad (11)$$

where $\tilde{h}^t$ is the predicted value of $h^t$ and $e^t$ describes $i$'th prediction-error in block $k$ related to the corresponding target pixel, $h^t$. Let's denote $\mathbf{E}^k$ as whole PEs in block $k$ and $e_{m_k}$ as the maximum PE in $\mathbf{E}^k$, $e_{m_k} = \max(|\mathbf{E}^k|)$. As discussed in [7], by choosing the least possible $n_k$ that satisfies

$$e_{m_k} < 2^{n_k}, \ 0 \leq n_k \leq 8 \quad (12)$$

we may bring out a label, $\ell_k$,

$$\begin{cases} \ell_k = 8 - n_k - 1 & n_k \neq 0 \\ \ell_k = 8 & n_k = 0 \end{cases} \quad (13)$$

that represents embedding capacity as much as $\ell_k$-bits may definitely be achieved by each pixel in block $k$. When $n_k = 7$, $\ell_k$ is zero, meaning there exists no room in the block $k$. Regarding $n_k = 8$, $\ell_k$ is "-1" that is taken zero as well. In [7], we prove that if $\ell_k$ more-significant-bits of a pixel are replaced with secret data, an authorized recipient can losslessly reconstruct the original pixel having $\ell_k$ and the predicted value of the pixel.

Regarding the number of pixels in each block, $q \times w$, the provided embedding capacity for the block $k$ is achieved by

$$C_{b_k} = (\ell_k) \times (q \times w) \quad (14)$$

Assuming the number of $K$ blocks, the total attained capacity is

$$C = \sum_{k=1}^{k=K} C_{b_k} \quad (15)$$

Note that, three bits are required to preserve each $\ell_k$. Therefore pure embedding capacity given by

$$C_p = C - 3 \times K \quad (16)$$

### B. Encryption

Content owner can encrypt the original content by stream cipher and an input secret key, $K_e$, as illustrated in [7]. Assuming target pixels, $\mathbf{H} = \{h^1, \dots, h^t, \dots h^T\}$, encrypted pixels are denoted by $\hat{\mathbf{H}} = \{\hat{h}^1, \dots, \hat{h}^t, \dots \hat{h}^T\}$. Furthermore, data hider can encrypt secret data by a standard encryption algorithm with an input secret key $K_d$.

### C. Embedding

Regarding pre-processing, labels $\mathcal{L} = \{\ell_1, \dots, \ell_k, \dots \ell_K\}$ are calculated and used to accommodate data bits by

$$[\![\hat{h}^t]\!] = \sum_{i'=1}^{\ell_k}(2^{8-i'} \times d^t_{(i'-1)}) + \sum_{i'=\ell_k+1}^{8}(\{\hat{h}^t\}_{8-i'} \times 2^{8-i'}) \quad (17)$$

where $\{\hat{h}^t\}_{i'}$ describes $i'$'th bit of the encrypted target pixel located in block $k$ and $d^t_{i'}$ demonstrates $i'$'th bit of to-be-embedded data may be either secret data or labels. Through (17), data is embedded in $\hat{h}^t$ that forms the marked pixel, $[\![\hat{h}^t]\!]$. In the equation, any $\sum_{i=n}^{m}$ with $n > m$ is to be considered as the empty sum, i.e. zero. Embedding may be done within two steps as discussed in [7]. In the first step, labels are embedded starting with the block provides the best embedding capacity and proceeding for next blocks. Address of the starting block and its label are stored in three first pixels of the image. The intensity of these three pixels and secret data are embedded in the second step.

In an alternative procedure to enhance security, information of labels and starting block may be encrypted using a third secret key, $K_s$, shared between data hider and content owner. At the recipient side, labels may be attained individually by either data hider or content owner using $K_s$. Having labels, data hider can extract secret data and content owner can reconstruct the original image separately by their own keys.

### D. Extracting data and recovering the original image

The first step for extracting data bits and reconstructing the original image is obtaining labels. Achieving labels is a hierarchical procedure that begins by attaining address of starting block and its label. They are obtained through information was stored in three first pixels of the image. Having

| Algorithm 1: Recovering $h^t$ located in the block k. |
|---|
| $\langle e^t \rangle = \langle h^t \rangle - \tilde{h}^t$ |
| **if** ($\ell_k == 8$) **or** ($\ell_k == 0$) **then** |
|    $h^t = \langle h^t \rangle$ |
| **else if** $|\langle e^t \rangle| < 2^{(8-\ell_k-1)}$ **then** |
|    $h^t = \langle h^t \rangle$ |
| **else if** $\langle e^t \rangle < 0$ |
|    $h^t = \langle h^t \rangle + 2^{8-\ell_k}$ |
| **else if** $\langle e^t \rangle > 0$ |
|    $h^t = \langle h^t \rangle - 2^{8-\ell_k}$ |
| **end if** |

address of starting block and its label, extracting is initiated to achieve labels of next blocks. Consequently, we can progressively extract more labels in a hierarchical procedure that results in obtaining the whole labels. They will be used to bring out secret data from the remaining blocks. Generally, extracting data from a marked encrypted pixel $[\![\hat{h}^t]\!]$, located in block $k$, is given by

$$\hat{d}_{i'}^t = \frac{\sum_{i'=1}^{\ell_k}(2^{8-i'} \times \{[\![\hat{h}^t]\!]\}_{8-i'})}{2^{8-\ell_k}} \quad (18)$$

where $\hat{d}_{i'}^t$ is $(i')$'th bit of the extracted data. Authorized user who has the $K_d$ can decrypt the data.

Having labels, reconstructing of the original image losslessly can be realized. The process begins by the decryption of $[\![\hat{h}^t]\!]$ into $[\![h^t]\!]$. First few rows or columns of the original image are restored just by decryption because they have been remained intact during embedding process. These restored pixels may be the neighboring pixels of a target pixel that are used to losslessly reconstruct the predicted value of the target pixel ($\tilde{h}^t$). In the following, we demonstrate two steps to recover original pixels. The first step is to achieve an intermediate value, $\langle h^t \rangle$, by

$$\langle h^t \rangle = \sum_{i'=1}^{8-\ell_k}(2^{i'-1} \times \{[\![h^t]\!]\}_{i'-1}) + \sum_{i'=9-\ell_k}^{8}(2^{i'-1} \times \{\tilde{h}^t\}_{i'-1}) \quad (19)$$

and in the second step original pixel is reconstructed using Algorithm1. The reconstructed pixels are employed to recover other target pixels in a hierarchical procedure.

IV. EXPERIMENTAL RESULT

We have conducted several experiments to confirm the performance of the proposed method in hiding capacity, lossless reconstruction and error-free data extraction. Twelve images, Splash, F16, House, Lena, Man, Raffia, Baboon, Bark, Straw, Grass, Tiffany and Airplane from the USC-SIPI data base, [47], and two images Crowd and Barbara from the miscellaneous database, [48], are employed as test gray scale images. We also investigate the performance of the different predictors and the proposed framework employing two entire databases, BOWS-2 [49] and BOSSbase, [50], and 850 images of UCID, [51]. Peak signal to noise ratio (PSNR) is used to evaluate lossless reconstruction of the original image and attained bits per pixel (bpp) is employed as a metric of embedding rate (ER). The below mentioned ER is pure ER attained after omitting overhead data.

In this section, first of all, the performance of the proposed predictors is evaluated. Secondly, embedding capacity is analyzed employing different predictors. Finally, the proposed scheme is compared with the prior arts on entire two data bases of BOSS-2 and BOSSbase and six standard test images to confirm the efficiency of our general framework.

*A. Performance of the proposed predictors*

Applying different predictors including proposed ones, we compute prediction-errors for 10,000 images of BOWS-2 and 856 images of UCID. The better predictor we choose the smaller PEs we achieve that results in the sharper histogram of the errors. The average histogram of the PEs for absolute values of 0, 1 and 2 is demonstrated in Fig. 4 for images of BOWS-2 and UCID. Note that, MED, GAP and proposed predictors are causal, while CB and LD predictors are non-causal. As described, CB predictor provides the best prediction and LD the worst. Because CB is non-causal, is not advantageous to be employed in our general framework. As shown, L predictor outperforms SGAP for both databases. By using BOWS-2, in comparison with GAP, L predictor attains sharper histogram of the PEs while it is much less complex. However for UCID images, GAP averagely makes better prediction than L predictor. As a result, L predictor can be as efficient as GAP and MED in prediction while being much less complex. Furthermore, applying adaptive L predictor in a block size of $16 \times 16$, denoted by AL16, we tangibly outperform conventional casual predictors, MED and GAP.

*B. Analysis of the embedding capacity*

The proposed algorithm is implemented using different predictors in three various sizes of blocks, $2 \times 4$, $3 \times 3$ and $4 \times 4$. The average ER is listed in Table II for test images.

For each block, a label of 3 bits is dedicated. Accordingly, the required overhead data bits are 97920, 86700 and 49152 for dividing the image into the non-overlapped blocks of sizes $2 \times 4$, $3 \times 3$ and $4 \times 4$, respectively. Two rows or columns of the image are discounted in the process of embedding. As tabulated, employing the block size $2 \times 4$ reaches more payload than others, generally. Meanwhile, smoother images such as F16 and Splash achieve the average ER higher than 3.5 bpp. For rougher ones such as Baboon, Bark, Straw, the average ER is less than 1.6 bpp and even less than 0.82 for Grass.

In comparison between the proposed predictors, the AL predictor results in more ER than the L predictor. Results will be superior if we apply AL in separated blocks of the image, namely, AL blocking. The smaller the block we choose, the more efficient the coefficient and consequently the greater the capacity we achieve while the overhead data is enlarged as well. We devote fifteen bits to store computed coefficient related to each block. As a compromise, an optimal size of a block may be adopted. Accordingly, AL blocking is done in sizes of $16 \times 16$, $32 \times 32$ and $64 \times 64$ that are denoted by AL16, AL32 and AL64 and consequently the required bits of overhead are 15360, 3840 and 960, respectively. As described in Table II, AL32 predictor leads more payload than AL16 and AL64 predictors, generally. Furthermore, AL32 predictor noticeably outperforms other predictors, GAP, MED, L and AL. In Grass, Straw, Barbara and Baboon images this improvement is more remarkable, e. g. for Straw image in using blocks of sizes $2 \times 4$ and 3 × 3, AL32 predictor almost achieves 0.4 bpp improvement over GAP. In comparison between L, GAP and MED predictors, the performance is different for various images. Accordingly, for House and Raffia images, MED predictor results in better ER than the GAP and L predictors while on average for all block sizes GAP is slightly better than L and L a little better than MED.

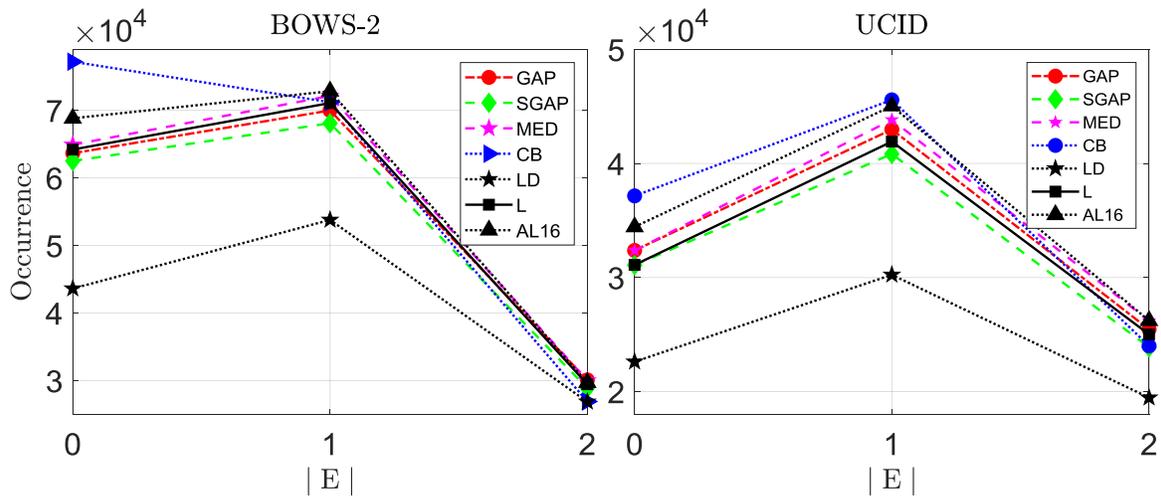

Fig. 4. Performance comparisons between the proposed predictors, L and AL16, and others on two databases BOWS-2 and UCID.

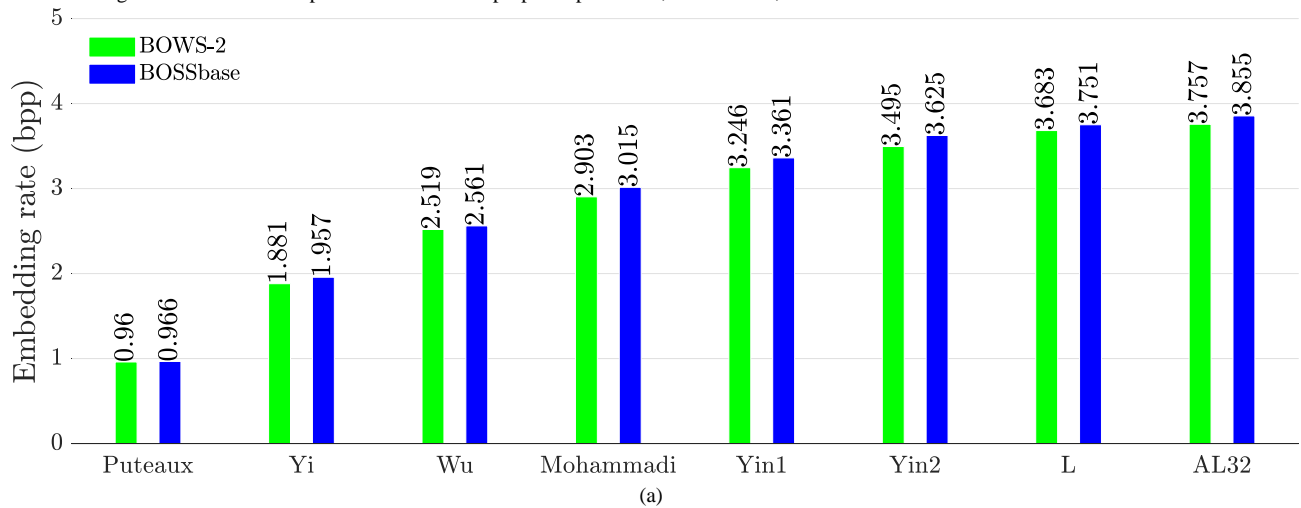

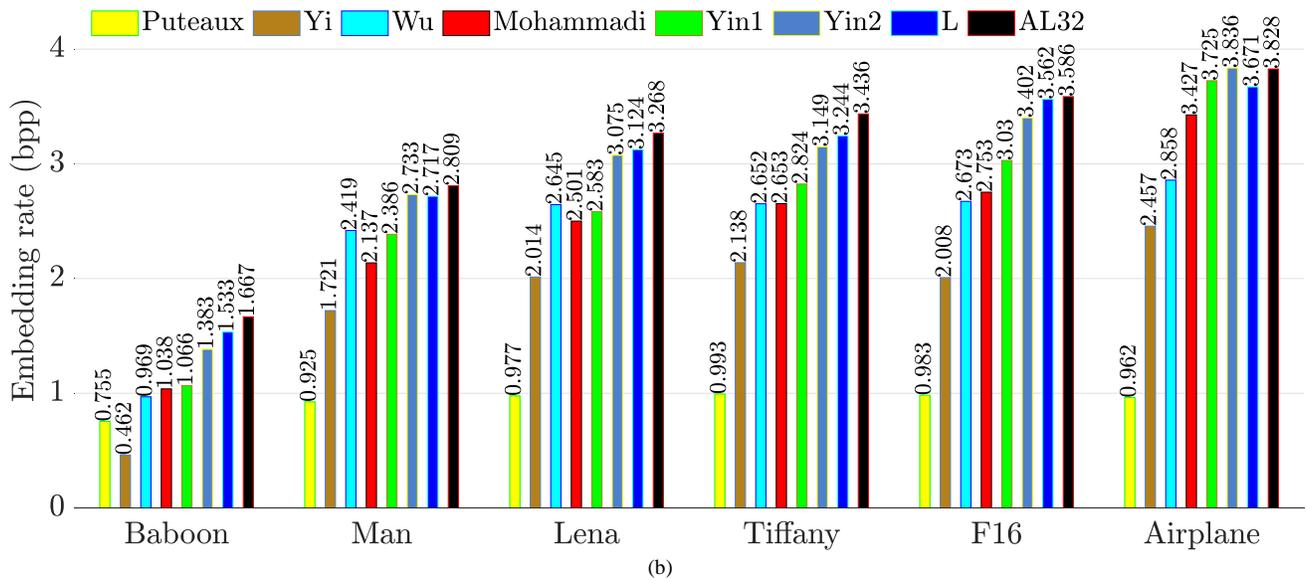

Fig. 5. Performance comparison between our general framework in adopting L and AL32 predictors and state of the art schemes, Puteaux [8], Yi [19], Wu [10], Mohammadi [7], Yin1 [9] and Yin2 [12] a) on two entire databases, BOWS-2 and BOSSbase, b) for standard test images, Baboon, Man, Lena, Tiffany, F16 and Airplane.

In more experiments, we employ BOSSbase, BOWS-2 and UCID data bases to more intuitively evaluate the performance of the proposed scheme in employing various predictors. The average ER is listed in Table III for various sizes of block. Generally, the results is better by using block size 2 × 4 rather than 3 × 3 and 4 × 4. For instance, by applying UCID, average ER through the block size 2 × 4 is 3.093 bpp which is almost 0.1 bpp better than using the block size 4 × 4.

In comparison between L, GAP and MED predictors, on average, L predictor slightly provides less ER. However, employing BOWS-2, L predictor reaches more ER than MED and GAP in all sizes of block. Since L predictor is much less complex than MED and GAP, the results confirm that proposed L predictor generally is more efficient than conventional causal predictors, GAP and MED.

Taking AL blocking, we significantly outperform the L predictor. For example, AL32 outperforms L predictor almost 0.1 bpp on average.

For all experiments, lossless reconstruction is fulfilled, i.e. PSNR = ∞, just by having $K_e$ and error-free data extraction is realized just by having $K_d$.

## C. Comparisons with prior arts

In Fig. 5, employing L and AL32 predictors, the proposed framework is compared with prior arts Puteaux [8], Yi [19], Wu [10], Mohammadi [7], Yin1 [9] and Yin2 [12]. In this comparison, the average ER for entire images of BOWS-2 and BOSSbase is calculated for prior arts and the proposed scheme. In our framework, the block size 2 × 4 and two predictors L and AL32 are employed. In Yi's scheme, [19], RDHEI is realized via VRBE while the proposed scheme and other discussed prior arts accomplish RDHEI by RRBE. In aforementioned schemes of RRBE, the main step of pre-processing is computing prediction-errors. Meanwhile, [7-10] and the proposed scheme bring out some labels from prediction-errors that are used after image encryption for embedding data. The labels, as overhead data, should be preserved in the encrypted image to make possible error-free data extraction and lossless reconstruction of the original image at the recipient side. In Yin1's scheme, [9], these labels are losslessly compressed to reduce overhead data. Their work may be somehow considered as a customized scheme of our general framework when GAP predictor and block size 1 × 1 are adopted. In this case, due to enlarging the number of blocks, which is equal to the number of target pixels, there exist huge overhead data bits that must be compressed by entropy coding to be compacted for conveying. However, employing larger size of the block we get rid of the need for entropy coding. Furthermore, through more efficient labeling procedure and proper predictors, our general framework outperforms Yin1's scheme more than 0.49 bpp on both data bases. Mohammadi's scheme, [7], can be considered as a specific plan of our general framework in which the LD predictor is chosen. Employing more efficient predictors in an innovative framework, we remarkably improve our previous work. In another scheme of Yin *et al*. [12], Yin2, PEs are obtained by MED predictor and losslessly compressed to release room. In ER, we surpass this scheme more than 0.22 bpp employing AL32 predictor. However, we suppose that content-owner is not enthusiastic to use lossless compression or entropy coding.

TABLE II EMBEDDING CAPACITY PROVIDED BY THE PROPOSED ALGORITHM FOR THE TEST IMAGES IN VARIOUS BLOCK SIZES.

| $E^K$ | Predictor | images | | | | | | | | | | | | | |
|---|---|---|---|---|---|---|---|---|---|---|---|---|---|---|---|
| | | Splash | F16 | Crowd | House | Lena | Barbara | Man | Raffia | Baboon | Bark | Straw | Grass | Average | |
| 2 × 4 | GAP | 3.743 | 3.581 | 3.439 | 3.279 | 3.219 | 2.64 | 2.757 | 2.247 | 1.532 | 1.543 | 1.261 | 0.774 | **2.501** | bpp |
| | MED | 3.681 | 3.509 | 3.354 | 3.303 | 3.079 | 2.511 | 2.639 | 2.293 | 1.449 | 1.469 | 1.066 | 0.753 | **2.426** | bpp |
| | L | 3.68 | 3.562 | 3.414 | 3.229 | 3.124 | 2.512 | 2.717 | 2.203 | 1.533 | 1.505 | 1.126 | 0.796 | **2.450** | bpp |
| | AL | 3.655 | 3.496 | 3.445 | 3.239 | 3.19 | 2.574 | 2.751 | 2.331 | 1.563 | 1.552 | 1.44 | 0.851 | **2.507** | bpp |
| | AL64 | 3.747 | 3.577 | 3.461 | 3.316 | 3.243 | 2.801 | 2.79 | 2.331 | 1.652 | 1.551 | 1.598 | 0.859 | **2.577** | bpp |
| | AL32 | 3.773 | 3.586 | 3.462 | 3.349 | 3.268 | 2.877 | 2.809 | 2.323 | 1.667 | 1.546 | 1.63 | 0.855 | **2.595** | bpp |
| | AL16 | 3.751 | 3.572 | 3.457 | 3.34 | 3.267 | 2.923 | 2.8 | 2.291 | 1.649 | 1.518 | 1.634 | 0.824 | **2.586** | bpp |
| | Average | **3.719** | **3.555** | **3.433** | **3.294** | **3.199** | **2.691** | **2.752** | **2.288** | **1.578** | **1.526** | **1.394** | **0.816** | | bpp |
| 3 × 3 | GAP | 3.72 | 3.558 | 3.413 | 3.25 | 3.208 | 2.623 | 2.758 | 2.165 | 1.516 | 1.519 | 1.245 | 0.764 | **2.478** | bpp |
| | MED | 3.659 | 3.49 | 3.328 | 3.287 | 3.071 | 2.492 | 2.641 | 2.248 | 1.436 | 1.451 | 1.061 | 0.743 | **2.409** | bpp |
| | L | 3.666 | 3.535 | 3.396 | 3.225 | 3.114 | 2.506 | 2.721 | 2.166 | 1.52 | 1.497 | 1.124 | 0.799 | **2.439** | bpp |
| | AL | 3.635 | 3.473 | 3.418 | 3.231 | 3.178 | 2.557 | 2.753 | 2.3 | 1.551 | 1.536 | 1.428 | 0.847 | **2.492** | bpp |
| | AL64 | 3.722 | 3.551 | 3.440 | 3.301 | 3.225 | 2.779 | 2.791 | 2.299 | 1.637 | 1.536 | 1.584 | 0.856 | **2.560** | bpp |
| | AL32 | 3.745 | 3.560 | 3.437 | 3.332 | 3.250 | 2.854 | 2.808 | 2.293 | 1.651 | 1.531 | 1.614 | 0.853 | **2.577** | bpp |
| | AL16 | 3.725 | 3.548 | 3.431 | 3.324 | 3.249 | 2.896 | 2.8 | 2.258 | 1.632 | 1.507 | 1.615 | 0.821 | **2.567** | bpp |
| | Average | **3.696** | **3.531** | **3.409** | **3.279** | **3.185** | **2.672** | **2.753** | **2.247** | **1.563** | **1.511** | **1.382** | **0.812** | | bpp |
| 4 × 4 | GAP | 3.684 | 3.507 | 3.303 | 3.207 | 3.172 | 2.603 | 2.677 | 2.045 | 1.464 | 1.447 | 1.195 | 0.683 | **2.416** | bpp |
| | MED | 3.619 | 3.442 | 3.218 | 3.233 | 3.012 | 2.469 | 2.565 | 2.087 | 1.389 | 1.367 | 0.987 | 0.666 | **2.338** | bpp |
| | L | 3.624 | 3.492 | 3.292 | 3.163 | 3.078 | 2.487 | 2.643 | 2.017 | 1.478 | 1.409 | 1.055 | 0.731 | **2.372** | bpp |
| | AL | 3.602 | 3.434 | 3.324 | 3.165 | 3.143 | 2.542 | 2.673 | 2.144 | 1.513 | 1.45 | 1.336 | 0.767 | **2.424** | bpp |
| | AL64 | 3.69 | 3.512 | 3.346 | 3.247 | 3.197 | 2.755 | 2.71 | 2.143 | 1.594 | 1.451 | 1.482 | 0.772 | **2.492** | bpp |
| | AL32 | 3.71 | 3.523 | 3.344 | 3.285 | 3.224 | 2.832 | 2.729 | 2.137 | 1.61 | 1.447 | 1.513 | 0.77 | **2.510** | bpp |
| | AL16 | 3.691 | 3.512 | 3.337 | 3.279 | 3.226 | 2.871 | 2.722 | 2.107 | 1.588 | 1.424 | 1.518 | 0.737 | **2.501** | bpp |
| | Average | **3.660** | **3.489** | **3.309** | **3.226** | **3.150** | **2.651** | **2.674** | **2.097** | **1.519** | **1.428** | **1.298** | **0.732** | | bpp |

TABLE III EMBEDDING CAPACITY PROVIDED BY THE PROPOSED FRAMEWORK FOR THE TEST IMAGES IN VARIOUS BLOCK SIZES.

| $E^K$ | Predictor | images | | | | |
|---|---|---|---|---|---|---|
| | | BOSS base | BOWS-2 | UCID | Average | |
| $2 \times 4$ | GAP | 3.772 | 3.680 | 3.096 | **3.516** | bpp |
| | MED | 3.765 | 3.677 | 3.077 | **3.506** | bpp |
| | L | 3.751 | 3.683 | 3.024 | **3.486** | bpp |
| | AL | 3.763 | 3.691 | 3.034 | **3.496** | bpp |
| | AL64 | 3.838 | 3.748 | 3.121 | **3.569** | bpp |
| | AL32 | 3.855 | 3.757 | 3.148 | **3.587** | bpp |
| | AL16 | 3.845 | 3.739 | 3.154 | **3.579** | bpp |
| | **Average** | **3.798** | **3.711** | **3.093** | | bpp |
| $3 \times 3$ | GAP | 3.745 | 3.651 | 3.057 | **3.484** | bpp |
| | MED | 3.743 | 3.654 | 3.043 | **3.480** | bpp |
| | L | 3.730 | 3.659 | 3 | **3.463** | bpp |
| | AL | 3.740 | 3.666 | 3.002 | **3.469** | bpp |
| | AL64 | 3.814 | 3.722 | 3.087 | **3.541** | bpp |
| | AL32 | 3.830 | 3.730 | 3.113 | **3.558** | bpp |
| | AL16 | 3.818 | 3.711 | 3.115 | **3.548** | bpp |
| | **Average** | **3.774** | **3.685** | **3.060** | | bpp |
| $4 \times 4$ | GAP | 3.696 | 3.593 | 2.990 | **3.426** | bpp |
| | MED | 3.691 | 3.594 | 2.974 | **3.420** | bpp |
| | L | 3.684 | 3.604 | 2.934 | **3.407** | bpp |
| | AL | 3.692 | 3.610 | 2.943 | **3.415** | bpp |
| | AL64 | 3.770 | 3.671 | 3.029 | **3.490** | bpp |
| | AL32 | 3.788 | 3.680 | 3.058 | **3.509** | bpp |
| | AL16 | 3.780 | 3.665 | 3.066 | **3.504** | bpp |
| | **Average** | **3.729** | **3.631** | **2.999** | | bpp |

In another experiment, the often-adopted test images Baboon, Man, Lena, Tiffany, F16 and Airplane are employed to evaluate our method compared with the prior arts. As shown in Fig. 5b, Yin2 surpasses other state of the art schemes for entire test images. In Airplane, there exist no noticeable difference of ER between our method in using AL32 and Yin2. As shown for other test images, our general framework improves Yin2's scheme. In Tiffany, this improvement is almost 0.3 bpp and in Baboon it is even more than twenty percent.

## V. CONCLUSION

In this paper, a general framework of RRBE schemes is presented to accommodate data in the encrypted image using advantages of different predictors including proposed ones. We evaluate performance of the proposed predictors in comparison with traditional ones that proves ours are more efficient. Generally, in using different predictors, we averagely achieve more than 3.4 bpp that confirms proposed scheme is a high capacity one. It is worth noting that, error-free data extraction and perfect reconstruction of the original image are realized in our framework for all experiments. Employing proper predictors, blocking of the PEs and more efficient procedure for labeling we surpass prior arts without using lossless compression. For examples, using block size $2 \times 4$ and AL32 predictor we averagely improve Yin's schemes more than 0.2 bpp on two entire data bases of BOWS-2 and BOSSbase. However, they employ lossless compression to release room. As a future work, we are willing to improve embedding capacity adopting smaller sizes of the block and lossless compression.